\documentclass[a4paper,fleqn]{cas-dc}

\usepackage[authoryear]{natbib}

\usepackage{subfigure}
\usepackage{graphicx}
\usepackage{amsmath}

\def\tsc#1{\csdef{#1}{\textsc{\lowercase{#1}}\xspace}}
\tsc{WGM}
\tsc{QE}

\begin{document}
\let\WriteBookmarks\relax
\def\floatpagepagefraction{1}
\def\textpagefraction{.001}

\shorttitle{}    

\shortauthors{Jiang}  

\title [mode = title]{Gradient Entropy (GradEn): The Two-dimensional Version of Slope Entropy for Image Analysis}  



%

\author[1]{Runze Jiang}[orcid=0000-0003-4106-4900]
\ead{23111517@bjtu.edu.cn}
\cormark[1]
\cortext[1]{Corresponding author}

\author[1]{Pengjian Shang}
\ead{pjshang@bjtu.edu.cn}

\author[2]{Du Shang}
\ead{dushang@bjtu.edu.cn}

\affiliation[1]{organization={School of Mathematics and Statistics}, 
	addressline={Beijing Jiaotong University},
	city={Beijing}, 
	country={China}}

\affiliation[2]{organization={School of Automation and intelligence}, 
	addressline={Beijing Jiaotong University},
	city={Beijing}, 
	country={China}}
















\begin{abstract}
Information theory and Shannon entropy are essential for quantifying irregularity in complex systems or signals. Recently, two-dimensional entropy methods, such as two-dimensional sample entropy, distribution entropy, and permutation entropy, have been proposed for analyzing 2D texture or image data. This paper introduces Gradient entropy (GradEn), an extension of slope entropy to 2D, which considers both symbolic patterns and amplitude information, enabling better feature extraction from image data. We evaluate GradEn with simulated data, including 2D colored noise, 2D mixed processes, and the logistic map. Results show the ability of GradEn to distinguish images with various characteristics while maintaining low computational cost. In classification tasks involving real-world datasets - including texture analysis, fault gear diagnosis, and railway corrugation detection - GradEn also exhibits significantly better performance than conventional 2D entropy methods. In conclusion, GradEn is an effective tool for image characterization, offering a novel approach for image processing and recognition.

\end{abstract}




\maketitle

\section{Introduction}
\label{sec1}

Entropy-based algorithms have emerged as fundamental tools for quantifying the irregularity and uncertainty inherent in time series analysis and complex signal characterization (\cite{bib1,bib2,bib3,bib4}). Originating from Shannon's information theory\cite{Shannon}, these methods operate by analyzing the probability distribution of distinct states within temporal data, thereby revealing hidden structural patterns in dynamical systems. Over recent decades, the progressive development of entropy quantification techniques has addressed diverse challenges in real world applications, with prominent methodologies including: approximate entropy\cite{ApEn}, sample entropy\cite{SampEn}, fuzzy entropy (\cite{FuzzEn}), permutation entropy (\cite{PerEn}), distribution entropy (\cite{DistrEn}) and dispersion entropy (\cite{DispEn}). Each methodology provides distinct mathematical frameworks for quantifying system complexity. The practical effectiveness of these entropy measures has been demonstrated across diverse application domains, including biological signal processing, financial market analysis, and industrial system monitoring, where they exhibit remarkable capabilities in differentiating complex signal patterns.

Recent years have witnessed a significant expansion of information theory and entropy-based methodologies into texture and image analysis. Building upon their success in one-dimensional signal processing, two-dimensional entropy measures have emerged as powerful tools for quantifying texture complexity by capturing the spatial irregularity and uncertainty inherent in image data. This methodological evolution has led to the development of specialized 2D entropy algorithms, including two dimensional sample entropy ($SampEn_{2d}$) (\cite{SampEn2D}), two dimensional distrbution entropy ($DistrEn_{2d}$) (\cite{DistrEn2D}), two dimensional permutation entropy ($PerEn_{2d}$) (\cite{PerEn2D}) and two dimensional dispersion entropy ($DispEn_{2d}$) (\cite{DispEn2D}). all of which have demonstrated efficacy in analyzing both synthetic textures and real-world images.

In this study, we introduce Gradient Entropy (GradEn), a novel two-dimensional extension of Slope Entropy (SloE) (\cite{SloE}), specifically designed for analyzing 2D datasets. Building on the strengths of its one-dimensional counterpart, GradEn uniquely combines both symbolic representations and amplitude variations at each spatial point. This integration enhances GradEn's discriminative power, allowing for more effective characterization of diverse image data and its classification applications. We begin by discussing the selection of parameters and the robustness of the GradEn method when applied to 2D white noise. Next, experimental validation of classification capability is carried out through systematic testing on simulated 2D datasets, which comprise colored noise and Mix processes. For logistic mapping, we employ the distance matrix from the recurrence plot method to convert the 1D time series into an image, enabling the identification of distinct images generated by different logistic parameters. Additionally, the computational cost is discussed to demonstrate the efficiency of GradEn. The method is also applied to real-world data: two real texture datasets are selected for classification tasks, and GradEn is used to detect and diagnose gear faults and railway corrugation in real-world signals after transforming these signals into matrix form. In both simulated and real-world experiments, the results of the proposed GradEn method are compared with other state-of-the-art 2D entropy algorithms, including $SampEn_{2d}$, $DispEn_{2d}$ and $PerEn_{2d}$. The classification results highlight the superiority and effectiveness of GradEn in distinguishing different types of textures and images, providing a novel perspective for analyzing the properties and structures of 2D data. 

The content of this paper is listed as follows: Section~\ref{sec2} provides a detailed explanation of the procedure for calculating GradEn. Section~\ref{sec3} presents the results of applying GradEn to simulated 2D data, demonstrating the effectiveness of the proposed method. In Section~\ref{sec4}, we use real-world texture datasets and signals for classification using GradEn. Finally, Section~\ref{sec5} concludes the paper.

\section{Gradient Entropy}
\label{sec2}

Slope entropy (\cite{SloE}) is based on calculating the slope between consecutive points in a time series. This method offers a unique approach by incorporating both the symbolic pattern and the amplitude information within the series. When applied to 2D images, the concept of slope can be extended and replaced by the gradient of pixels in various directions.

\begin{figure}
	\centering
	\includegraphics[width=3.5in]{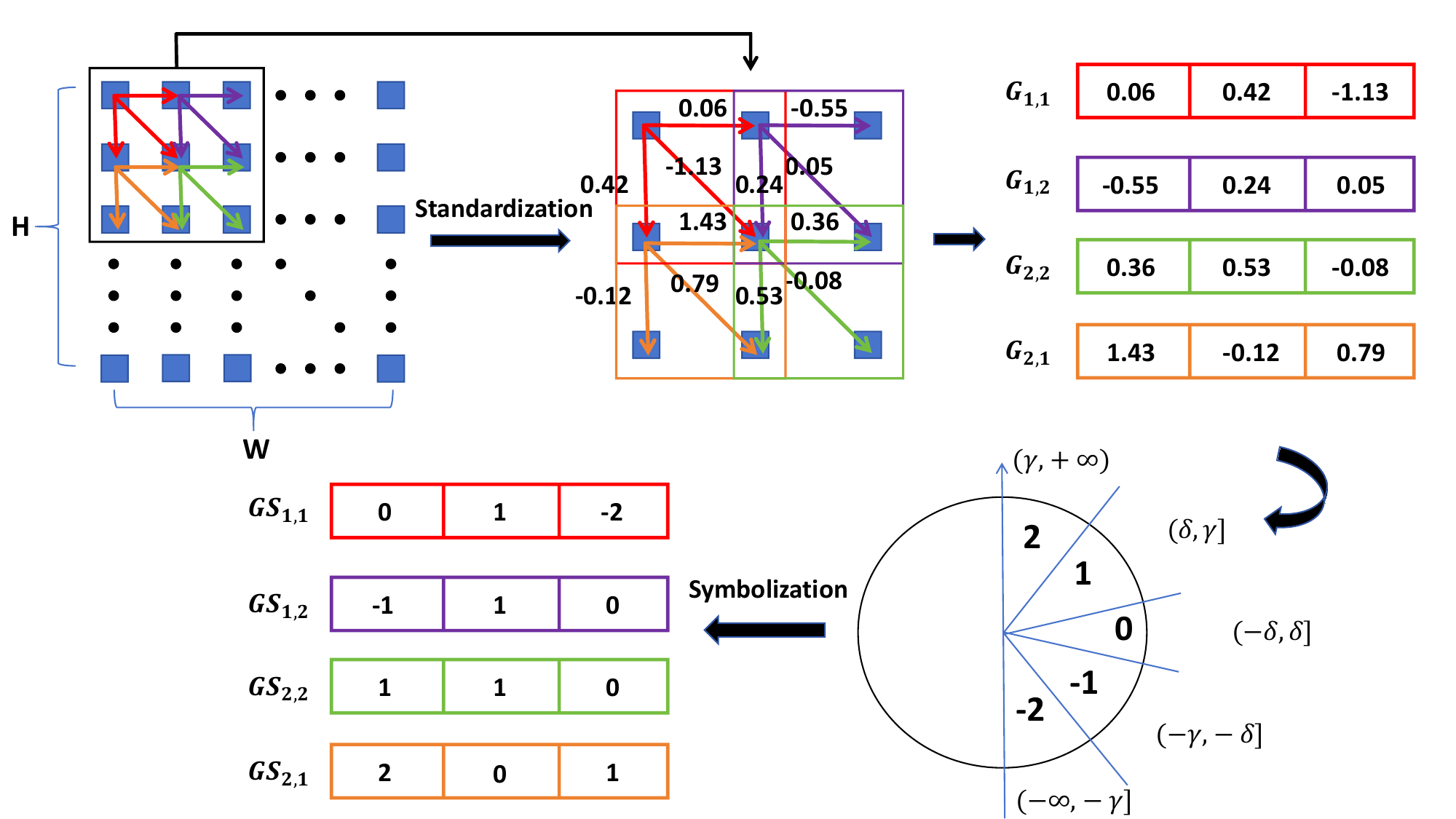}
	\caption{\label{fig1} The sketch map of GradEn symbolization algorithm.}
\end{figure} 

Consider an image $X=\{x_{i,j}\}_{i=1,2,...,H}^{j=1,2,...,W}$ with the size equals to $H \times W$. The sketch map of GradEn symbolization algorithm is shown in Fig.~\ref{fig1} and the procedure is listed as follows: 

The first step is standardizing the gradients present in the original image. Specifically, for each pixel $x_{i,j}$, where $i=1,2,...,H-1, j=1,2,...,W-1$, gradients in three directions --- horizontal ($G_{i,j}^h=x_{i,j+1}-x_{i,j}$), vertical ($G_{i,j}^v=x_{i+1,j}-x_{i,j}$) and diagonal ($G_{i,j}^d=x_{i+1,j+1}-x_{i,j}$), are computed. These gradients are then used to form a three-dimensional vector $G_{i,j}=(G_{i,j}^h,G_{i,j}^v,G_{i,j}^d)$. Next, the gradient vectors from all pixels are aggregated into a single vector $G_{x}=(G_{1,1},G_{1,2},...,G_{1,j-1},G_{2,1},...,G_{H-1,W-1})$. A z-score normalization is applied to $G_{x}$, obtaining the standardized data vector $GS_{x}=(GS_{1,1},GS_{1,2},...,GS_{1,j-1},GS_{2,1},...,GS_{H-1,W-1})$. It is important to note that each element $GS_{i,j}$ represents the standardized three-dimensional gradient of a 2*2 pixel block in the image.

The second step involves transforming each element $GS_{i,j}, i=1,2,...,H-1, j=1,2,...,W-1$ into symbols drawn from the set $(-2,-1,0,1,2)$. Similar to the SloE method, two thresholds, $\delta$ and $\gamma$ must be determined (the selection of these thresholds is discussed in Section 3). Next, each element of $GS_{i,j}$ is assigned a corresponding symbol, resulting in a new symbolic vector $GS_{i,j}^s$. For a given numeric value $a$, the following transformations are applied: 

$\bullet$ If $a\leq-\gamma$, assign the symbol -2. 

$\bullet$ If $-\gamma<a\leq-\delta$, transform it into -1.

$\bullet$ If $-\delta<a\leq\delta$, transform it into 0. 

$\bullet$ If $\delta<a\leq\gamma$, transform it into 1. 

$\bullet$ If $a>\gamma$, transform it into 2 .     

Finally, a series of three dimensional symbolic vectors $GS_{i,j}^s$ are obtained. There are a total of $5^3=125$ possible patterns, denoted as $\pi_{k},k=1,2,...,125$. The frequency of each pattern in the image is then calculated, with the corresponding formulation described as follows.

\begin{equation}
	\label{eq1}
	\resizebox{0.48\textwidth}{!}{
		$p(\pi_{k})=\frac{\|\,(i,j):1\leq i\leq H-1, 1\leq j\leq W-1\, , GS_{i,j}^s=type\,\{\pi_{k}\}\,\|}{(H-1)(W-1)}$}
\end{equation} 
where $\| \cdot \|$ represents the total number of all eligible elements. Then, the GradEn of image X is calculated by

\begin{equation}
	\label{eq2}
	GranEn(X)=\frac{-\sum_{i=1}^{5^3} p(\pi_{i}) \log{p(\pi_{i})}}{log{5^3}}.
\end{equation} 

\section{Experiments in simuated data}  
\label{sec3} 

\subsection{Parameter Analysis}
\label{sec3a}
\begin{figure}
	\centering
	\includegraphics[width=3in]{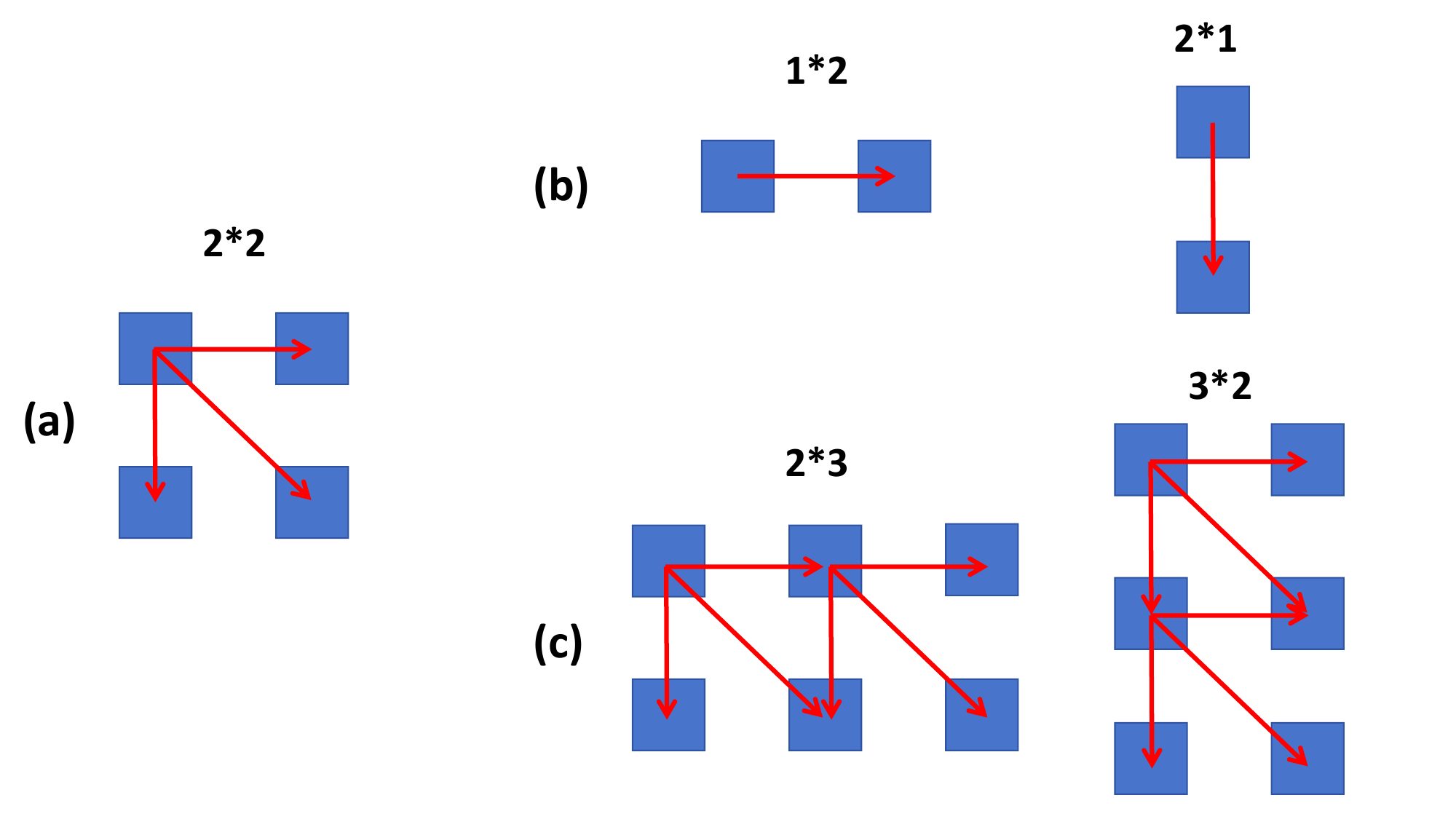}
	\caption{\label{fig2} Three types of pixel size of GradEn algorithm.}
\end{figure}

First of all, the selection of pixel block dimensions ($m*n$) in GradEn requires analysis, similar to the determination of embedding dimension $m$ in SloE. While the block size could theoretically be arbitrary, practical implementation must consider the structural characteristics of natural images. Smaller configurations such as $1*2$ or $2*1$ blocks yield only $5^1=5$ distinct gradient patterns(see Fig.~\ref{fig2}(b)), which proves insufficient for effective information characterization. Conversely, larger configurations like $2*3$ or $3*2$ blocks generate $5^6=15,625$ potential patterns (see Fig.~\ref{fig2}(c)), creating dual challenges of computational complexity and noise sensitivity. Through evaluation, we identify the $2*2$ configuration as an optimal compromise: it produces $5^3=125$ distinctive patterns - a quantity that maintains discriminative power while ensuring computational efficiency and noise robustness. 

\begin{figure}
	\centering
	\includegraphics[width=3.5in]{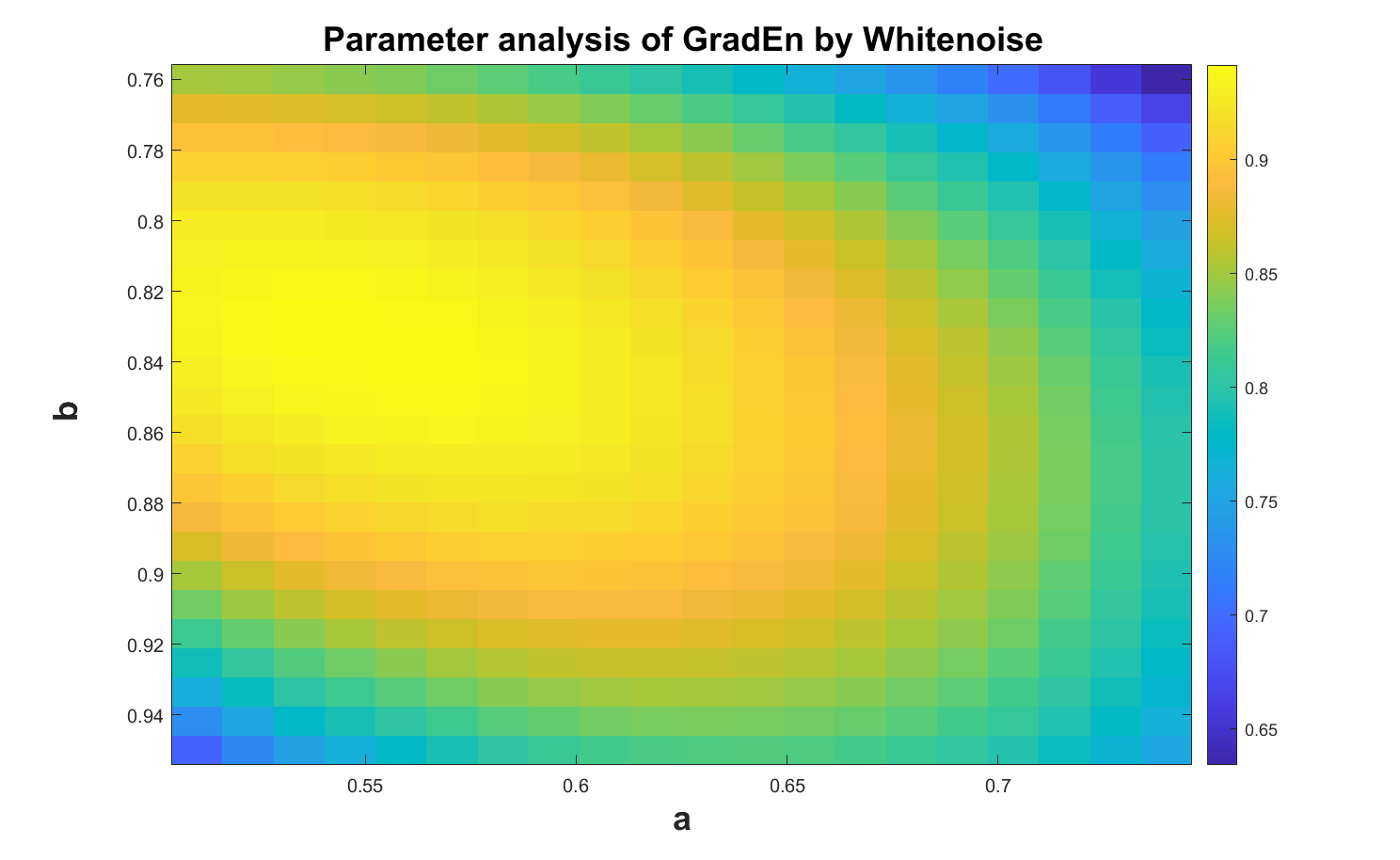}
	\caption{\label{fig3} The GradEn value of 2D Whitenoise with the change of $a$ and $b$.}
\end{figure}

Then, the thresholds $\delta$ and $\gamma$ should be selected carefully. Due to the z-score standardization, the range of $\delta$ is constrainted to $(\mu_{0.5},\mu_{0.75})$ and $\gamma$ falls within $(\mu_{0.75},\mu_{1})$, where $\mu_{\alpha}$ represents the $\alpha$ quantile of the normal distribution. Let $\delta=\mu_{a}$ and $\gamma=\mu_{b}$, a 100*100 2D whitenoise image with the size of 100*100 is then generated, and the GradEn values for $a$ changing from 0.51 to 0.74 and $b$ changing from 0.76 to 0.95 with 0.01 precision increments, are shown in Fig.~\ref{fig3}. Given the high degree of irregularity in the white noise data, the entropy value is expected to be relatively large. Therefore, the parameter values in the yellow region  ($a\in$ [0.53,0.6], $b\in$ [0.8,0.86]) in Fig.~\ref{fig3} are recommended. In this paper, $a=0.55$ and $b=0.8$ as the thresholds. 

\subsection{Colored noise classification}  
\label{sec3b}
\begin{figure}
	\centering
	\includegraphics[width=3.5in]{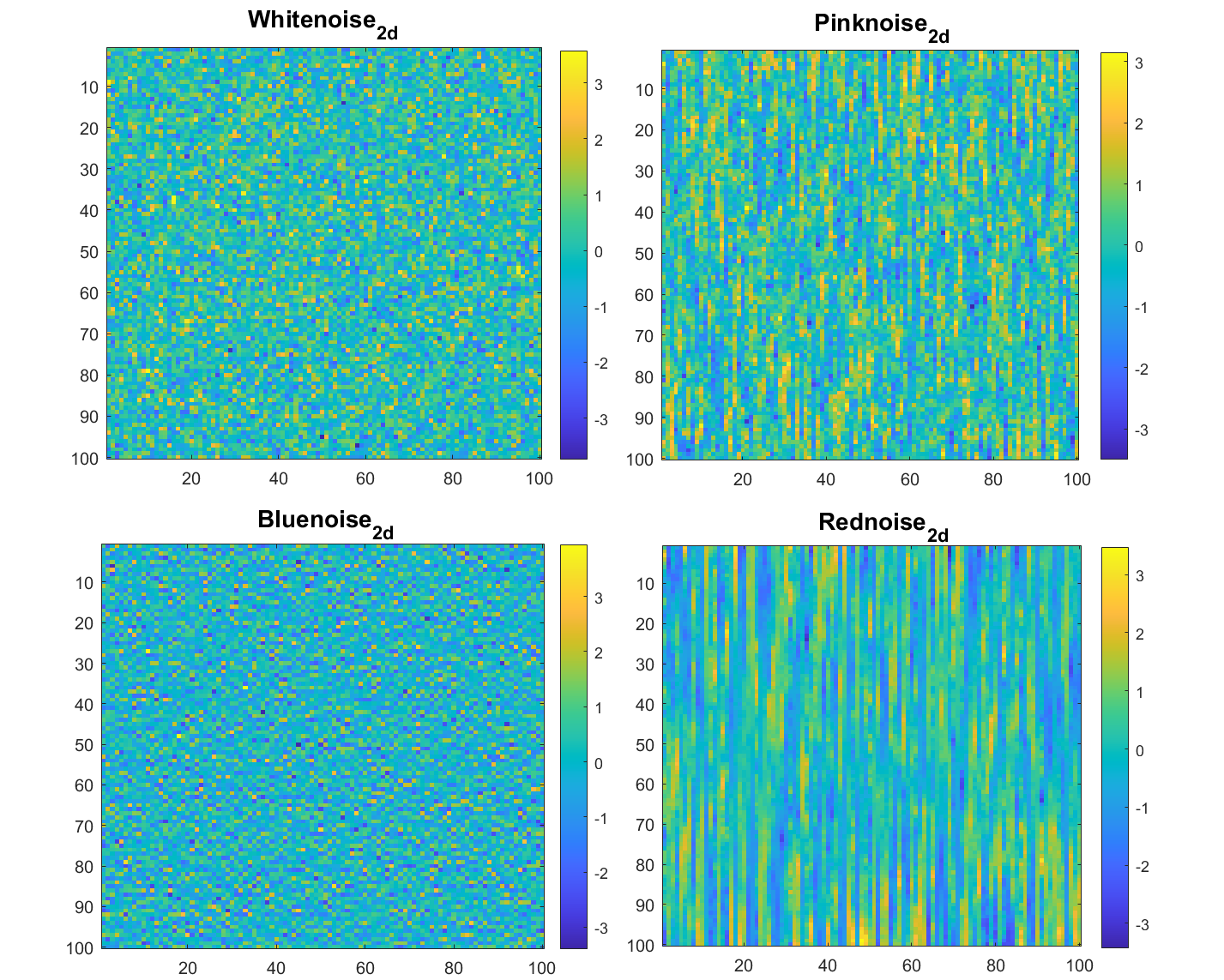}
	\caption{\label{fig4} The example of four types of 2D noises with the size equals to 100*100.}
\end{figure}

\begin{figure}
	\centering
	\includegraphics[width=3in]{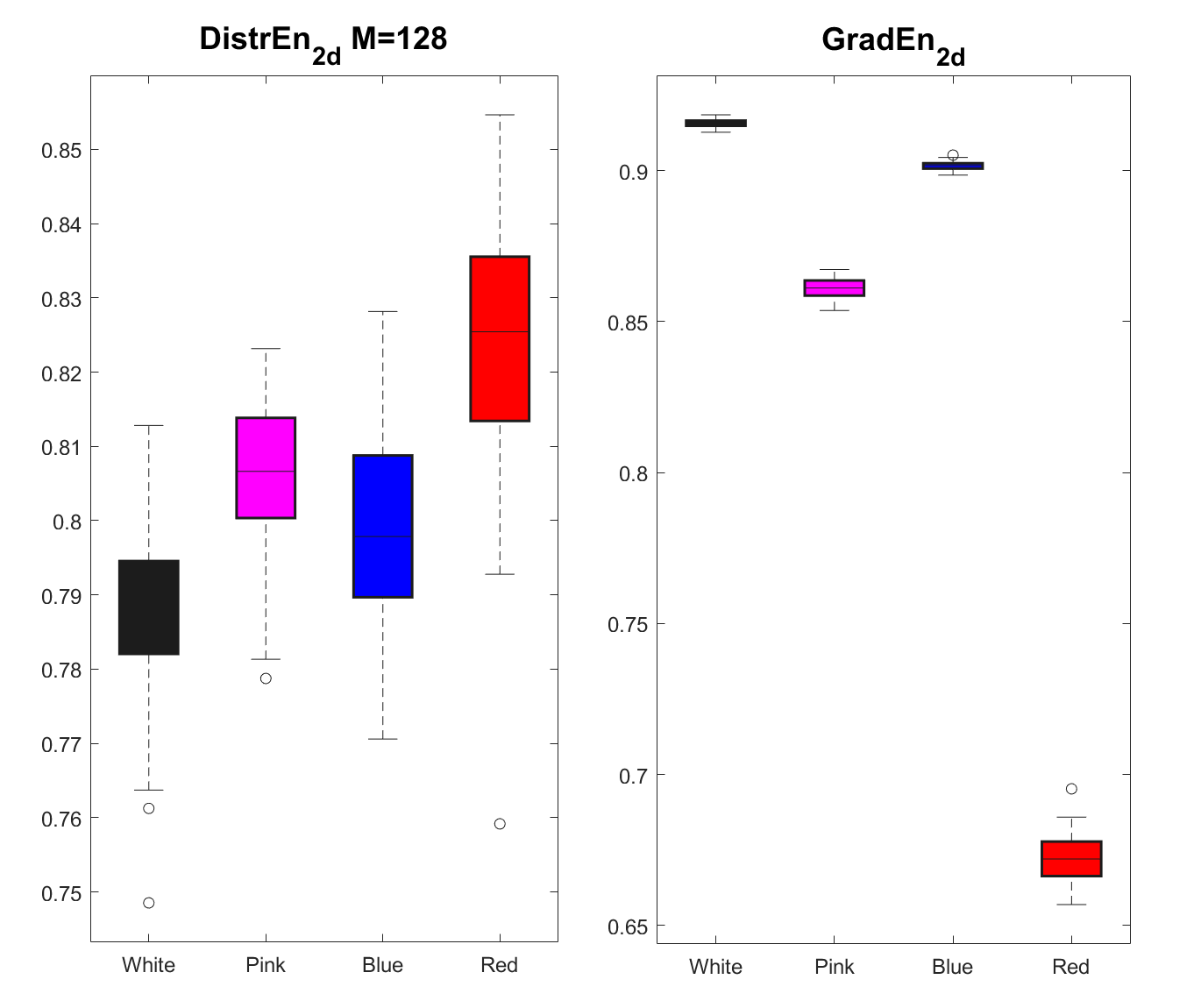}
	\caption{\label{fig5} The boxplot results of $DistrEn_{2d}$ (left) and $GradEn$ (right) of four type noises.}
\end{figure}

In this section, we examine the classification performance of GradEn using different types of 2D colored noise, including 2D white noise, 2D pink noise, 2D blue noise, and 2D red noise. The 2D noise images are generated by reshaping 1D noise sequences of length $N$ into $m*n$ matrices. For this experiment, each image is sized at 100x100, and 50 samples are generated for each type of noise. Examples of the four types of 2D noise are shown in Fig.~\ref{fig4}. Subsequently, boxplots for the four noise groups, calculated using both $DistrEn_{2d}$ and $GradEn$, are presented in Fig.~\ref{fig5}. The parameter $M$ is 128 in this experiment. As observed, while $DispEn_{2d}$ can somewhat distinguish between different noise types, the differences are not significant, particularly for pink noise and blue noise. In contrast, GradEn effectively distinguishes all four types of noise without any confusion, demonstrating its superior ability to differentiate between various image types.  

\subsection{Robust Analysis}
\label{sec3c}
\begin{figure}
	\centering
	\includegraphics[width=3.5in]{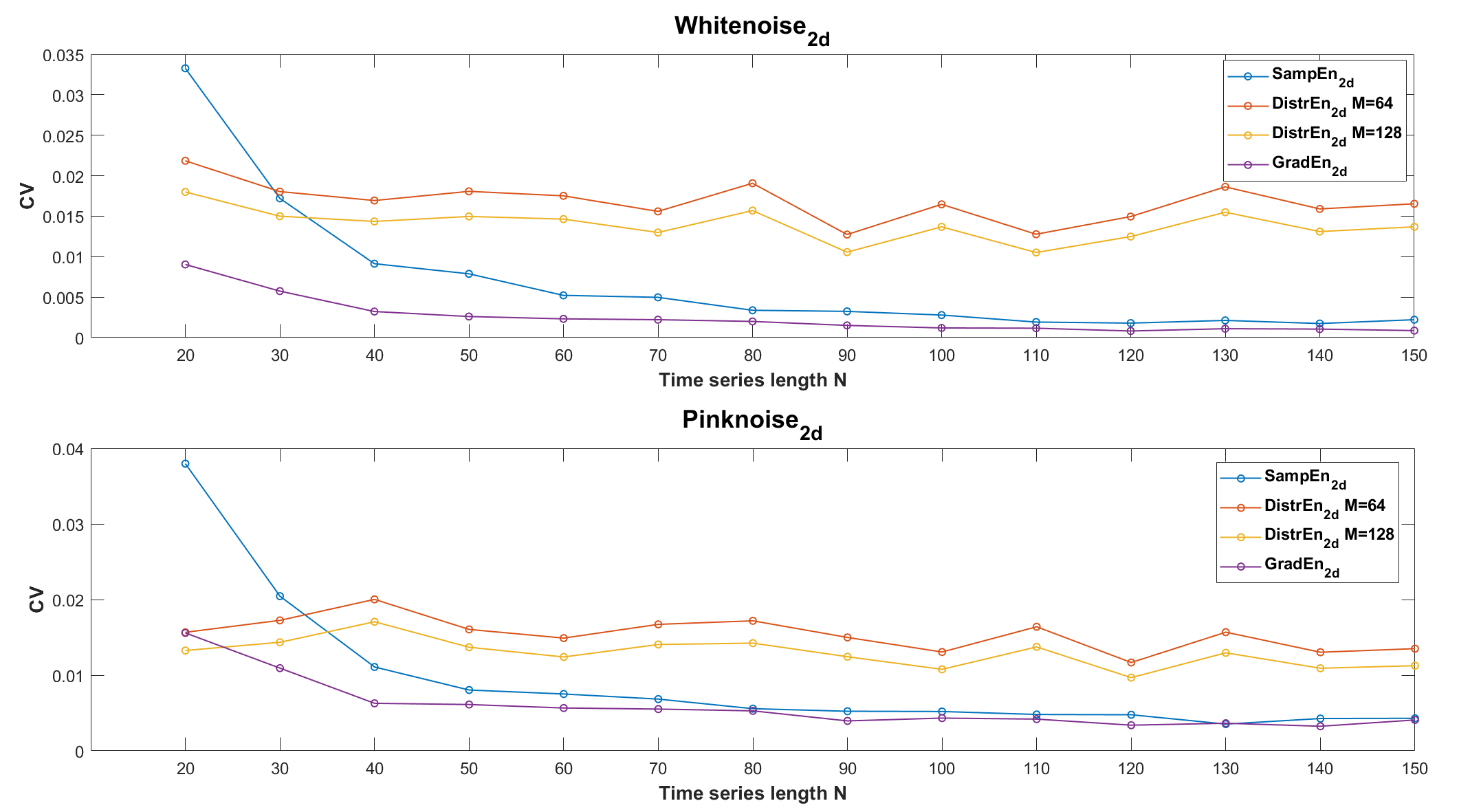}
	\caption{\label{fig6} The CV values of $SampEn_{2d}$, $DistrEn_{2d}$ and $GradEn$ applying 2D whitenoise and pinknoise under different image size.}
\end{figure}

The robustness of an algorithm is a critical performance indicator. In this section, we evaluate the robustness of the $GradEn$ method by comparing it with $SampEn_{2d}$ and $DistrEn_{2d}$ under varying conditions, including 2D noises of different lengths and mixed processes. We begin by generating 2D white noise and 2D pink noise samples in image sizes ranging from 20*20 to 150*150 , with a step size of 10. For each size, 100 noise samples are randomly generated. The coefficient of variation (CV) for each image size is calculated and presented in Fig.~\ref{fig6}. As shown, the CV values for $GradEn$ are consistently lower than those of $SampEn_{2d}$ and $DistrEn_{2d}$ across all image sizes, demonstrating the superior robustness of $GradEn$ under various conditions. 

\begin{figure}
	\centering
	\includegraphics[width=3.5in]{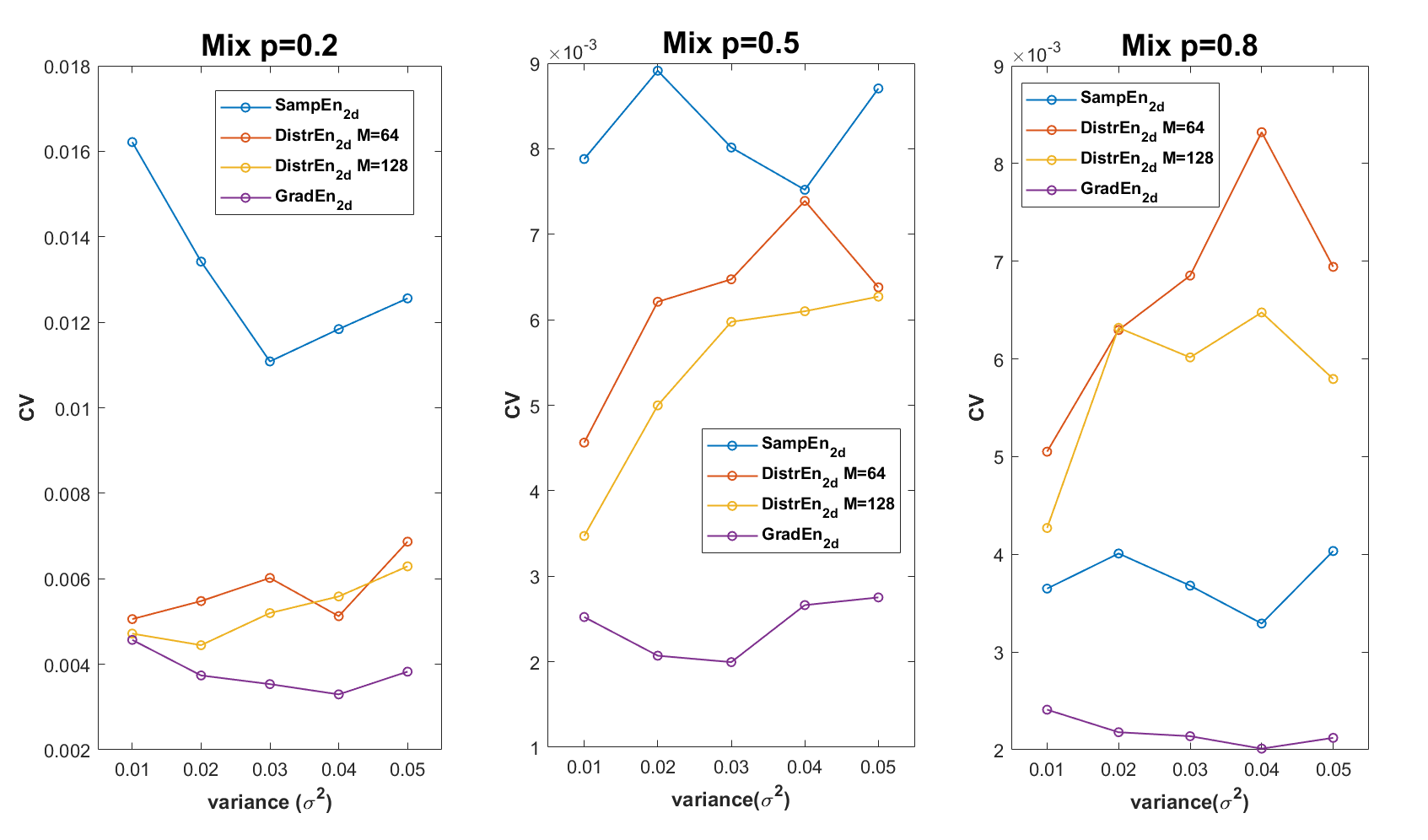}
	\caption{\label{fig7} The CV values of $SampEn_{2d}$, $DistrEn_{2d}$ and $GradEn$ by adding 2D whitenoise to Mix process with parameter $p$ equals to 0.2, 0.5, and 0.8.}
\end{figure}

Next, the robustness to noise is evaluated by applying the $MIX_{2d}$ process with different parameter values of $p$. The formulation of the $MIX_{2d}$ process is as follows:

\begin{equation}
	MIX_{2D}(p)_{i,j}=(1-Z_{i,j})X_{i,j}+Z_{i,j}Y_{i,j}
\end{equation}
where $X_{i,j}=sin(\frac{2 \pi i}{12})+sin(\frac{2 \pi j}{12})$ represents a periodic sine image, while $Y=Y_{i,j}$ denotes an image where all points are randomly generated from a uniform distribution within the range of [$-\sqrt{3},\sqrt{3}$]. $Z$ is a two-point distribution variable, where the probability of 1 is $p$ and the probability of 0 is $1-p$. As $p$ varies from 0 to 1, the characteristics of the mixed-process image shift from periodic to stochastic. In this section, we use the mixed process with parameters set to 0.2, 0.5 and 0.8, respectively. White noise with variance ranging from 0.01 to 0.05 is added to each of these three scenarios to calculate the coefficient of variation (CV) for each group. The comparison results, presented in Fig.~\ref{fig7}, show that $GradEn$ consistently yields the lowest CV values, highlighting the robustness of the GradEn method to noise.

\subsection{Logistic map}
\label{sec3d}
The logistic map is a widely used chaotic model due to its ability to exhibit a variety of nonlinear behaviors as the parameter changes\cite{logistic}. The formulation of the logistic map is given by:

\begin{equation}
	x_{i+1}=ax_{i}(1-x_{i})
\end{equation} 
where $a$ is the control parameter that determines the degree of periodicity or chaos in the sequence. To assess the ability of GradEn to distinguish between different states in the logistic map, we transform the logistic sequence into a distance matrix, similar to the preprocessing algorithm used in recurrence plots. Given the sequence $X_{i},i=1,2,...,N$, first perform phase space reconstruction with an embedding dimension $m$, resulting in a series of vectors $x(i)=(X_{i}, X_{i+1}, ..., X_{i+m-1})$ for $i=1,2,...,N-m+1$. Next, construct the distance matrix $D=D_{i,j}$, where $i=1,2,...,N-m+1$ and $j=1,2,...,N-m+1$, with each element defined as $D_{i,j}=dist(x(i), x(j))$. The Euclidean distance is used in this algorithm. Once the matrix $D$ is constructed, it is then processed using the 2D entropy method to calculate the results.

\begin{figure}
	\centering
	\includegraphics[width=3.6in]{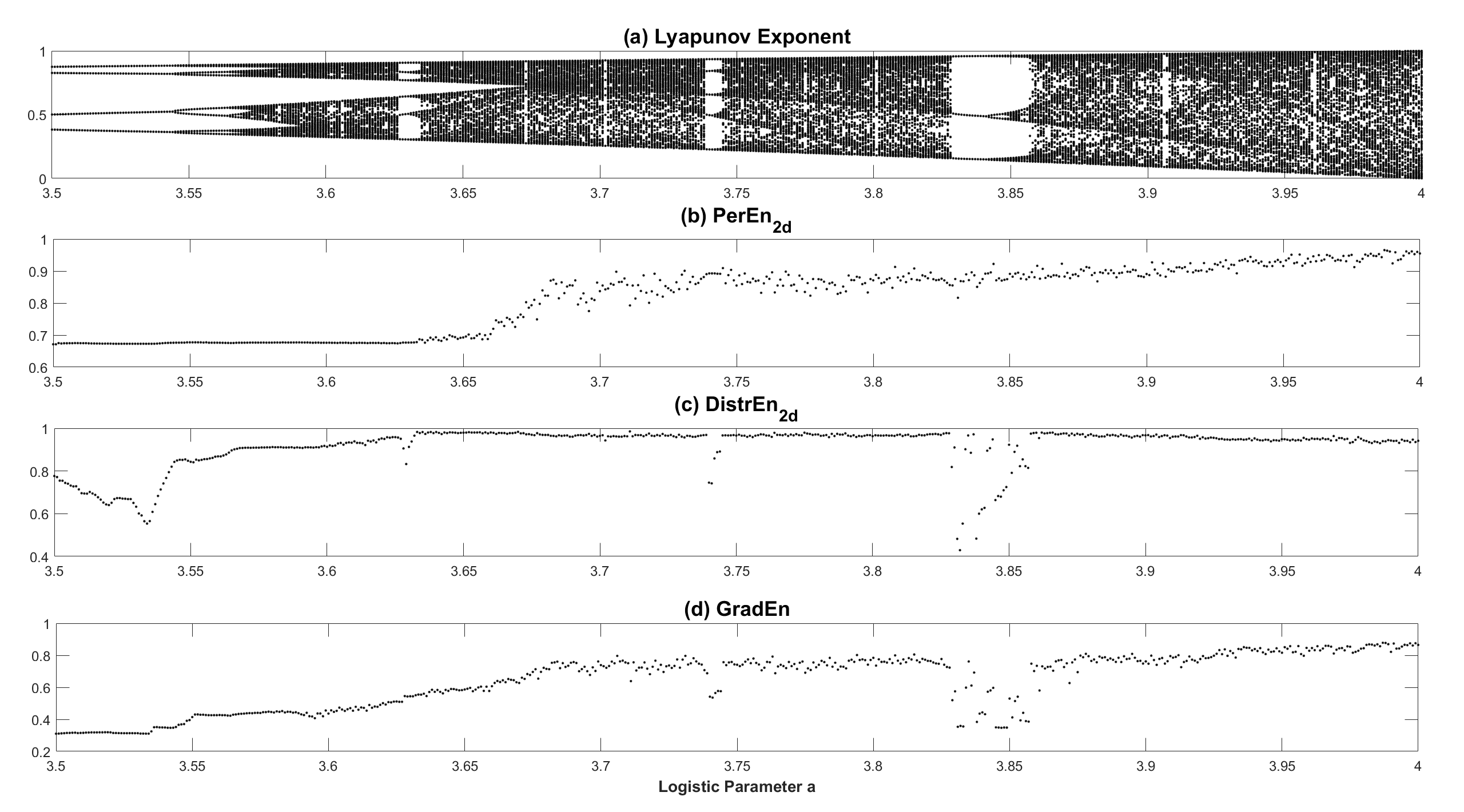}
	\caption{\label{fig8} The logistic map experiment results of $PerEn_{2d}$, $DistrEn_{2d}$ and $GradEn$ with the change of parameter $a$. (a) bifurcation graph. (b) $PerEn_{2d}$. (c) $DispEn_{2d}$. (d) $GradEn$.}
\end{figure}

The parameter $a$ is gradually varying from 3.5 to 4 with step equals to 0.01 in this experiment. The bifurcation graph, entropy results of $PerEn_{2d}$, $DistrEn_{2d}$ and $GradEn$ of different $a$ are presented in Fig.~\ref{fig8}. $m=3$ and $N=150$ is selected, the pixel block size of $PerEn_{2d}$ and $DistrEn_{2d}$ (M=128) are both 2*2. It can be concluded that $PerEn_{2d}$ can not distinguish periodic and chaotic sequence after $a>3.65$. Although $DispEn_{2d}$ can detect the periodic sequence from chaotic, it can not present the increase chaotic degree as $a$ go from 3.5 to 4. In comparison, $GradEn$ not only distinguish the periodic and chaotic sequences successfully, but also perform the increasing tendency with the change of $a$. This result proves the effectiveness of $GradEn$ in characterizing the property of different images. 

\subsection{Computation cost}
\label{sec3e}
In this section, the computational cost of the proposed $GradEn$ method is evaluated by comparing it with $SampEn_{2d}$ and $DistrEn_{2d}$. The 2D white noise images are generated with different sizes (40*40, 80*80, 120*120, 160*160), and the corresponding computation times under various conditions are presented in Table.~\ref{tab1}. The experiments are conducted using a 16-GB RAM system with Matlab R2021a. The results show that $SampEn_{2d}$ requires significantly more time to compute the entropy value of an image, especially when the image size exceeds  100*100. In contrast, $DistrEn_{2d}$ is more efficient than $SampEn_{2d}$. Among the three methods, $GradEn$ consistently demonstrates the shortest computation time across all conditions, with only a minimal increase in time as the image size grows. Therefore, the proposed $GradEn$ method proves to be an effective algorithm for quickly extracting features from images.

\begin{table}
	\centering
	\caption{The Hedges' g effect size results for the four methods across 15 EEG channels }
	\resizebox{0.8\linewidth}{!}{\begin{tabular}{cc|rrrr}
			\toprule
			\multicolumn{2}{c}{\multirow{2}{*}{\textbf{Entropy Measure}}} & 
			\multicolumn{4}{c}{\textbf{Image Size}} \\
			\cmidrule{3-6} \multicolumn{2}{c}{} &
			\multicolumn{1}{c}{40*40} & \multicolumn{1}{c}{80*80} &
			\multicolumn{1}{c}{120*120} & \multicolumn{1}{c}{160*160} 
			\\
			
			\midrule
			
			\multicolumn{2}{c}{\textbf{$SampEn_{2d} (m=1)$}} &  \multicolumn{1}{c}{3.033s} & \multicolumn{1}{c}{50.717s} & \multicolumn{1}{c}{262.28s} & \multicolumn{1}{c}{853.151s}   \\
			\cmidrule{1-2} 
			
			\multicolumn{2}{c}{\textbf{$SampEn_{2d} (m=2)$}} &  \multicolumn{1}{c}{3.229s} & \multicolumn{1}{c}{57.656s} & \multicolumn{1}{c}{301.41s} & \multicolumn{1}{c}{977.323s}   \\
			\cmidrule{1-2} 
			
			\multicolumn{2}{c}{\textbf{$SampEn_{2d} (m=3)$}} &  \multicolumn{1}{c}{2.989s} & \multicolumn{1}{c}{56.654s} & \multicolumn{1}{c}{300.31s} & \multicolumn{1}{c}{988.506s}   \\
			\cmidrule{1-2} 
			
			\multicolumn{2}{c}{\textbf{$DistrEn_{2d} (m=1)$}} &  \multicolumn{1}{c}{0.079s} & \multicolumn{1}{c}{0.088s} & \multicolumn{1}{c}{0.232s} & \multicolumn{1}{c}{1.195s}   \\
			\cmidrule{1-2} 
			
			\multicolumn{2}{c}{\textbf{$DistrEn_{2d} (m=2)$}} &  \multicolumn{1}{c}{0.072s} & \multicolumn{1}{c}{0.088s} & \multicolumn{1}{c}{0.257s} & \multicolumn{1}{c}{0.794s}   \\
			\cmidrule{1-2} 
			
			\multicolumn{2}{c}{\textbf{$DistrEn_{2d} (m=3)$}} &  \multicolumn{1}{c}{0.067s} & \multicolumn{1}{c}{0.086s} & \multicolumn{1}{c}{0.285s} & \multicolumn{1}{c}{1.229s}   \\
			\cmidrule{1-2} 
			
			\multicolumn{2}{c}{\textbf{$GradEn$}} &  \multicolumn{1}{c}{0.028s} & \multicolumn{1}{c}{0.082s} & \multicolumn{1}{c}{0.184s} & \multicolumn{1}{c}{0.431s}   \\

			\bottomrule
			\label{tab1}
	\end{tabular}}
\end{table}

\section{Experiments in real-world data}
\label{sec4}
\subsection{Texture Classification}   
\label{sec4a}
\begin{figure}
	\centering
	\includegraphics[width=3.5in]{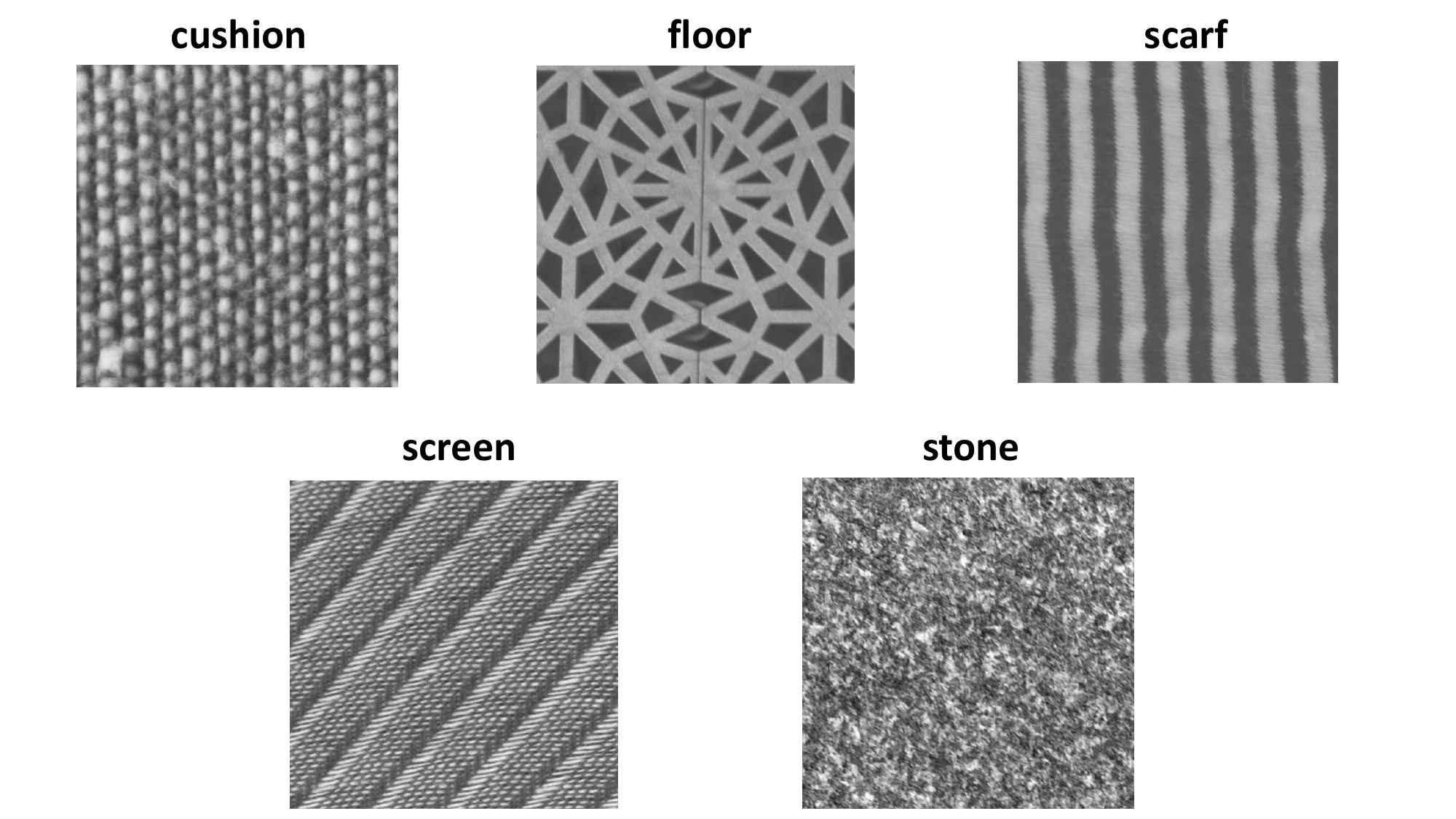}
	\caption{\label{fig9} The example of five type of textures in Kylberg texture dataset. }
\end{figure}

\begin{figure}
	\centering
	\includegraphics[width=3.5in]{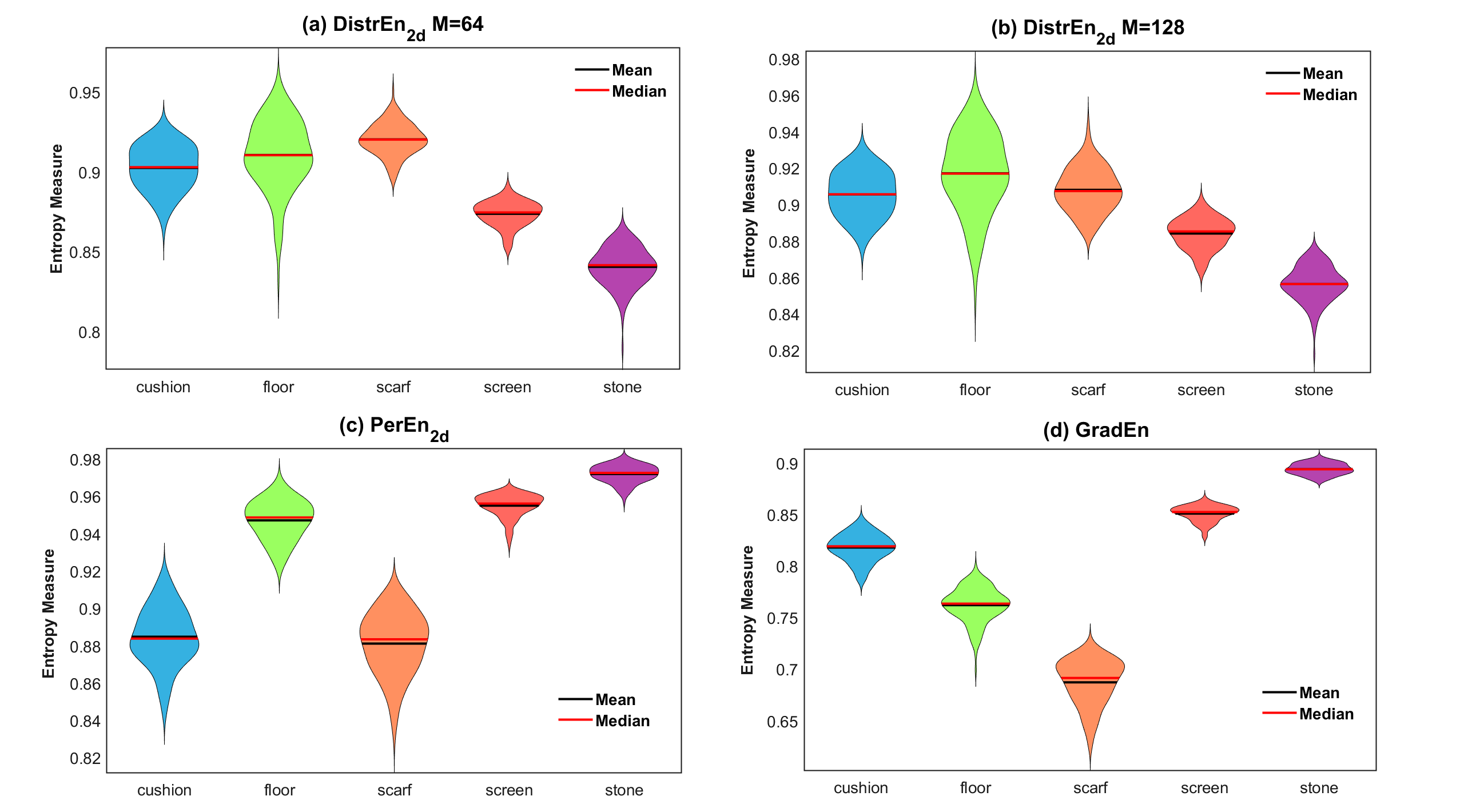}
	\caption{\label{fig10} The boxplot results of Kylberg texture dataset classification. (a) $DistrEn_{2d}$ (M=64). (b) $DistrEn_{2d}$ (M=128). (c) $PerEn_{2d}$. (d) $GradEn$.}
\end{figure}

Texture analysis is a key research area in image processing, focused on extracting the characteristics of various patterns within images. In this section, two texture datasets are used to evaluate the classification performance of the proposed $GradEn$ method. We begin by conducting experiments on the Kylberg texture dataset (\cite{Kylberg}), which includes five selected texture categories: cushion, floor, scarf, screen, and stone. Each category contains 160 samples, and examples of the images are shown in Fig.~\ref{fig9}. To standardize image size and ensure computational efficiency, all images are downsampled to 128×128. The boxplot results for $DistrEn_{2d}$ ($M=64$ and $M=128$), $PerEn_{2d}$, and $GradEn$ are presented in Fig.~\ref{fig10}. As shown in Fig.~\ref{fig10}(a) and (b), $DistrEn_{2d}$ fails to effectively distinguish between the five texture types, particularly for cushion, floor, and scarf. Similarly, $PerEn_{2d}$ ([Fig. 10(c)]) struggles to differentiate between cushion vs. scarf and floor vs. screen. In contrast, $GradEn$ demonstrates a clear ability to differentiate all five textures, as shown in [Fig. 10(d)], highlighting the superiority of the $GradEn$ method.  

\begin{figure}
	\centering
	\includegraphics[width=3.5in]{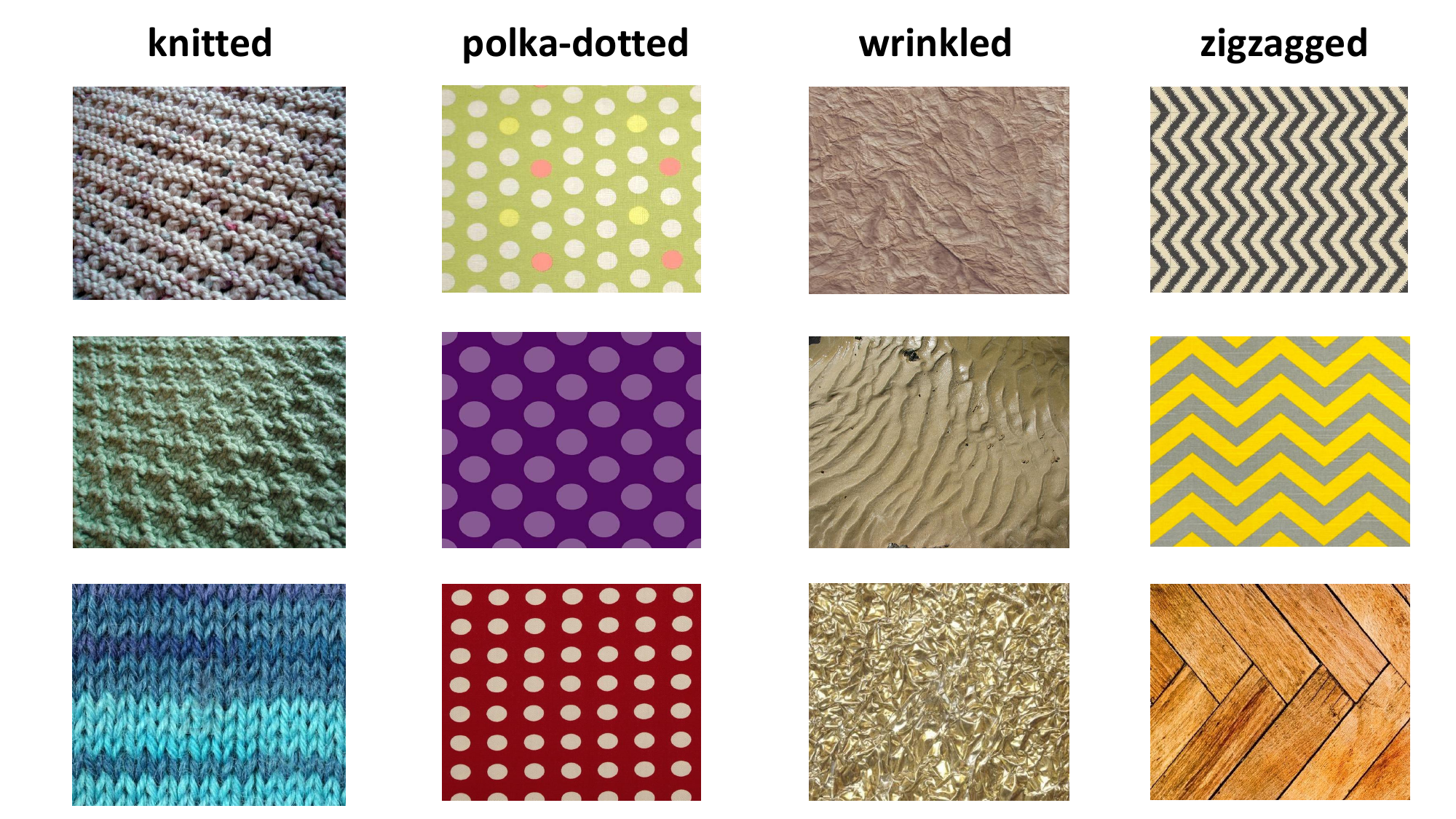}
	\caption{\label{fig11} The example of five type of textures in Describable texture dataset. }
\end{figure}

\begin{figure}
	\centering
	\includegraphics[width=3.5in]{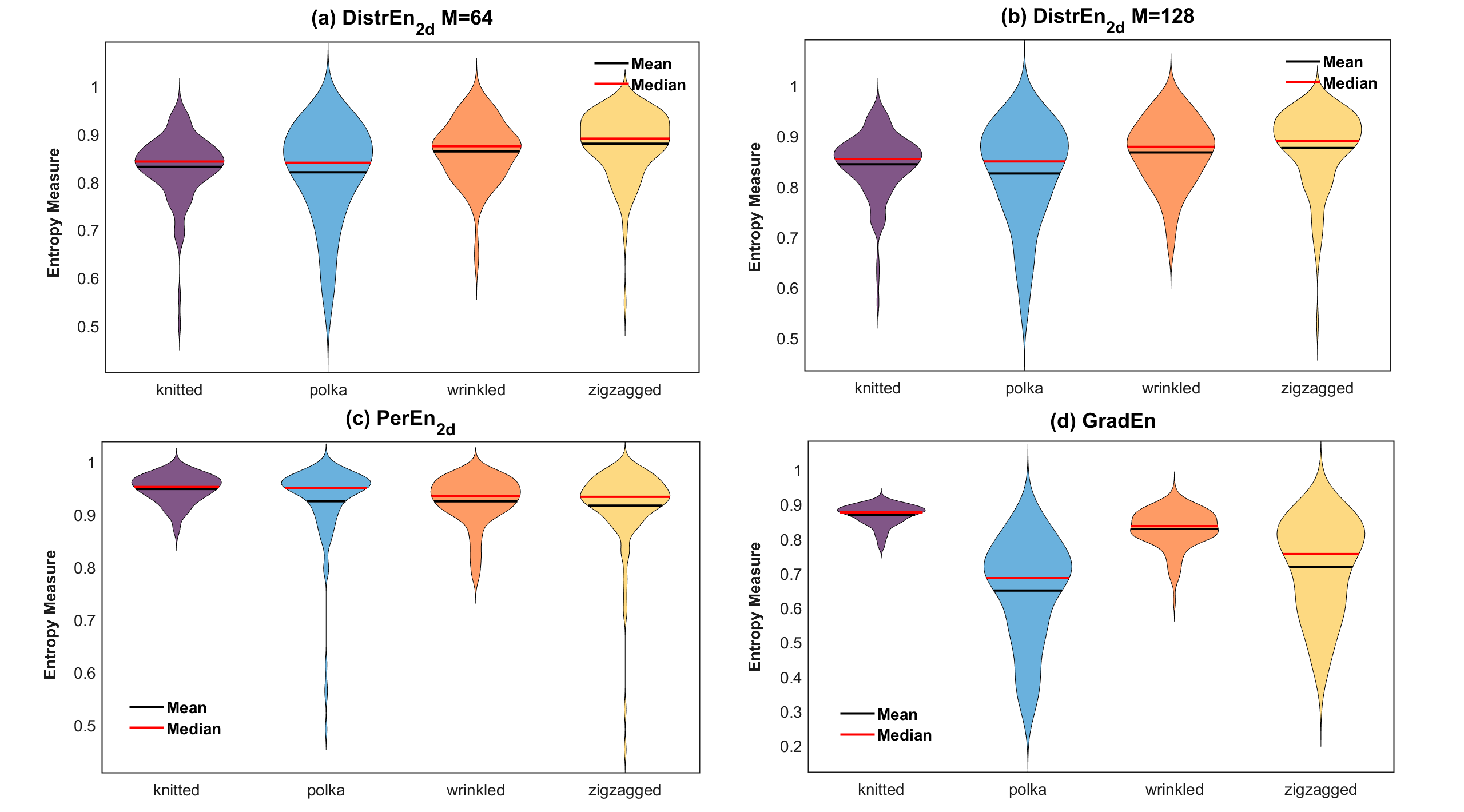}
	\caption{\label{fig12} The boxplot results of Describable texture dataset classification. (a) $DistrEn_{2d}$ (M=64). (b) $DistrEn_{2d}$ (M=128). (c) $PerEn_{2d}$. (d) $GradEn$.}
\end{figure}

Additionally, the Describable Texture Dataset (DTD) (\cite{DTD}) is used for the classification task. Four texture types—knitted, polka, wrinkled, and zigzagged—are selected, with 120 samples for each type and the examples are shown in Fig.~\ref{fig11}. All images are grayscaled processed by taking the means of RGB values and also downsampled to 128×128 for consistency. The boxplot results for the four entropy algorithms are presented in Fig.~\ref{fig12}. It is evident that only $GradEn$ is able to distinguish between the four texture groups to some extent, while the other three methods fail to differentiate the texture types effectively. In conclusion, both two experiments showcase the effectiveness of $GradEn$ for detecting the characteristic of texture images.  

\subsection{Gear fault diagnosis}
\label{sec4b}
\begin{figure}
	\centering
	\includegraphics[width=3.5in]{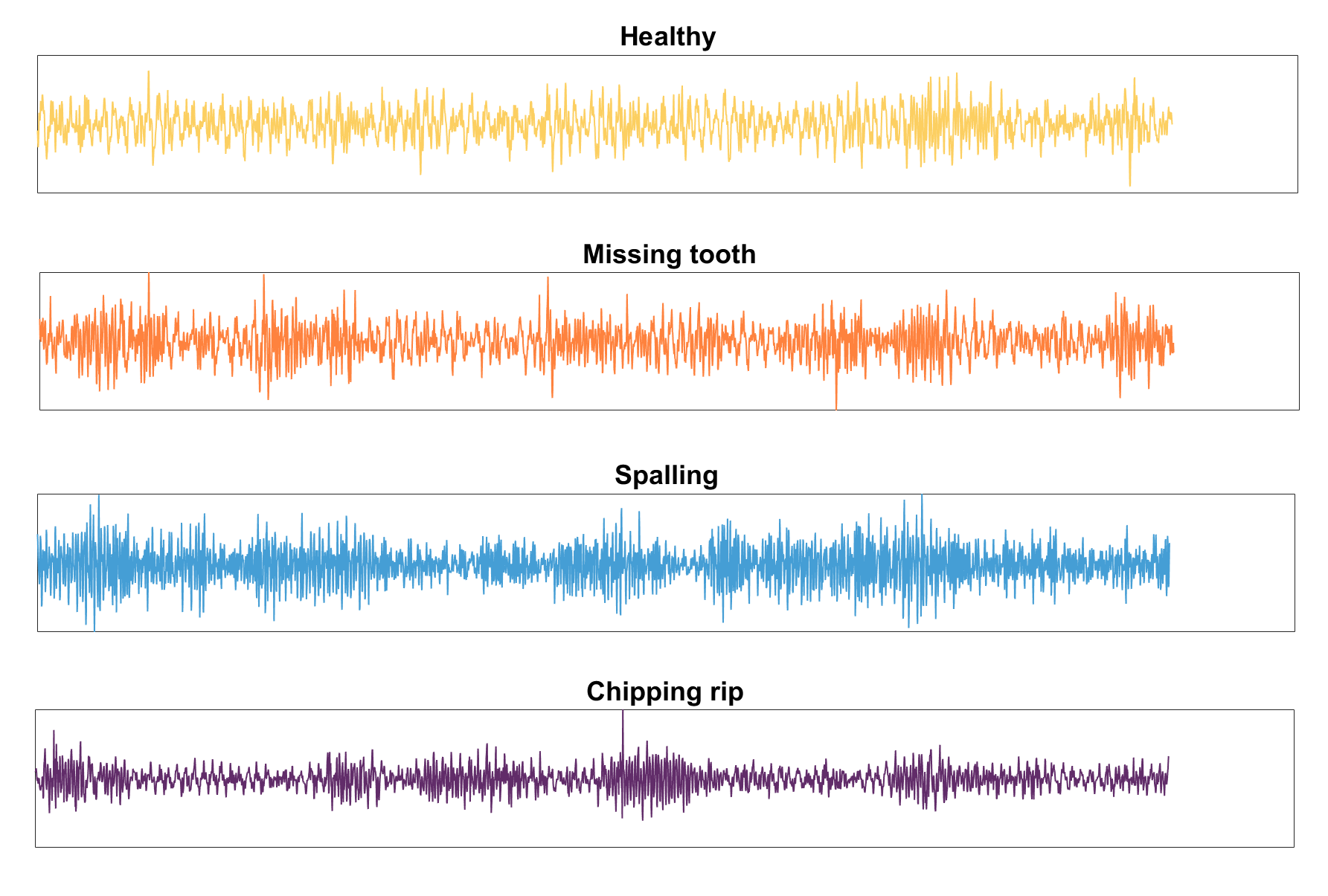}
	\caption{\label{fig13} The example of four types of fault gear signals.}
\end{figure}

\begin{figure}
	\centering
	\includegraphics[width=3.5in]{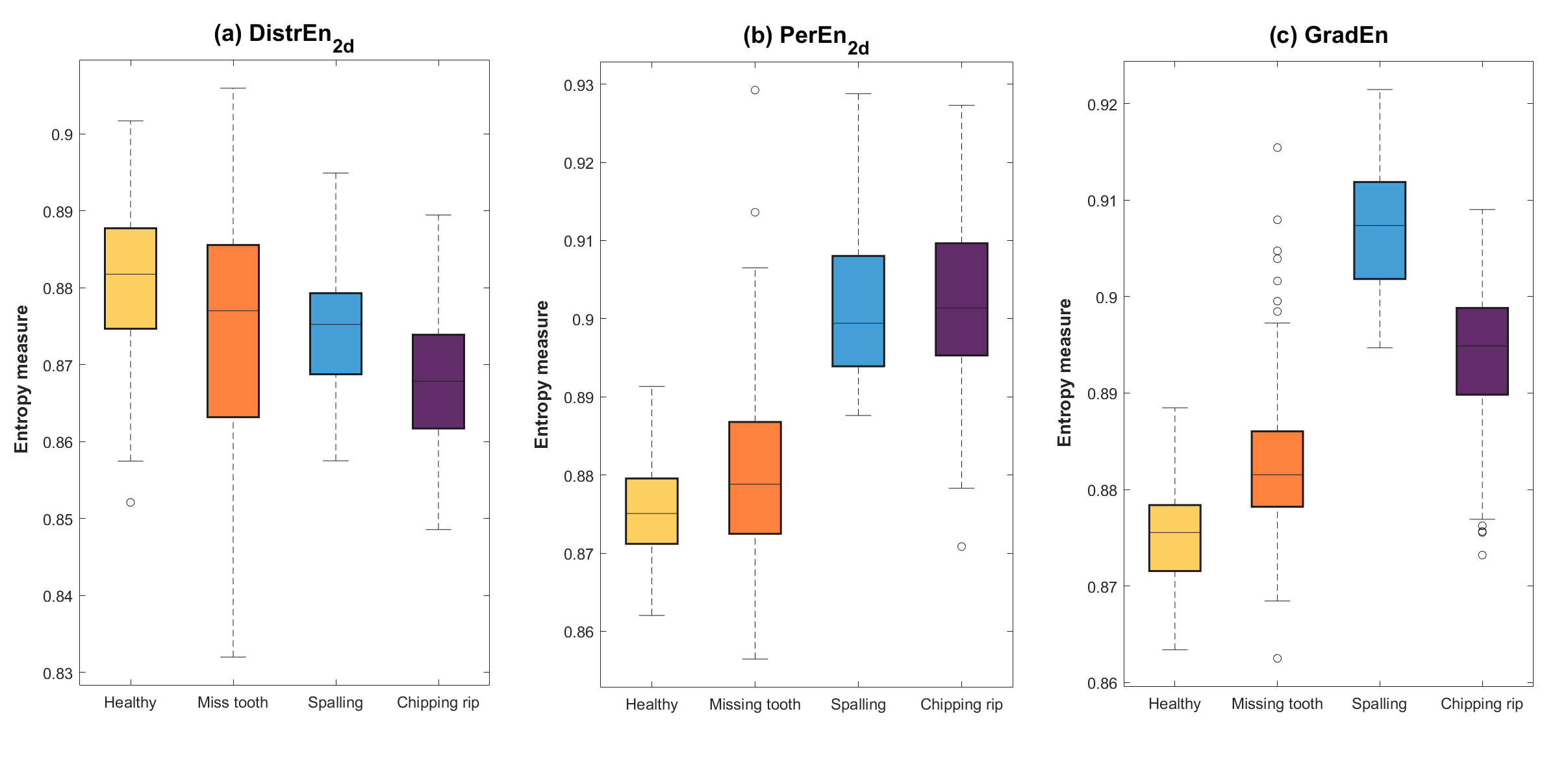}
	\caption{\label{fig14} The boxplot results of University of Connecticut gear fault dataset classification. (a) $DistrEn_{2d}$. (b) $PerEn_{2d}$. (c) $GradEn$.}
\end{figure}

Mechanical fault diagnosis plays a crucial role in industrial engineering, and numerous methods have been proposed by researchers in recent years (\cite{fault1,fault2,fault3,fault4}). In this section, gear fault data from the University of Connecticut are used for classification. The vibration signals are collected from a benchmark two-stage gearbox with replaceable gears. These signals are measured using an accelerometer and recorded through a dSPACE system, with a sampling frequency of 20 kHz. Four fault conditions—healthy, missing tooth, spalling, and chipping rip—are selected for the classification task (The examples are shown in Fig.~\ref{fig13}). Each fault type consists of 104 samples. The first 150 points of each sample are used to construct the distance matrix with an embedding dimension of $m=3$. The boxplot results for classifying the four gear faults are shown in Fig.~\ref{fig14}. As seen in the figure, $DistrEn_{2d}$ fails to effectively distinguish between the different fault types. $PerEn_{2d}$ struggles to differentiate between healthy vs. missing tooth and spalling vs. chipping rip. In contrast, only the $GradEn$ method is able to distinguish all four gear fault types based on entropy values, demonstrating its capability to characterize images generated by engineering signals. 

\subsection{Railway Corrugation Diagnosis}
\label{sec4c}
\begin{figure}
	\centering
	\includegraphics[width=3.5in]{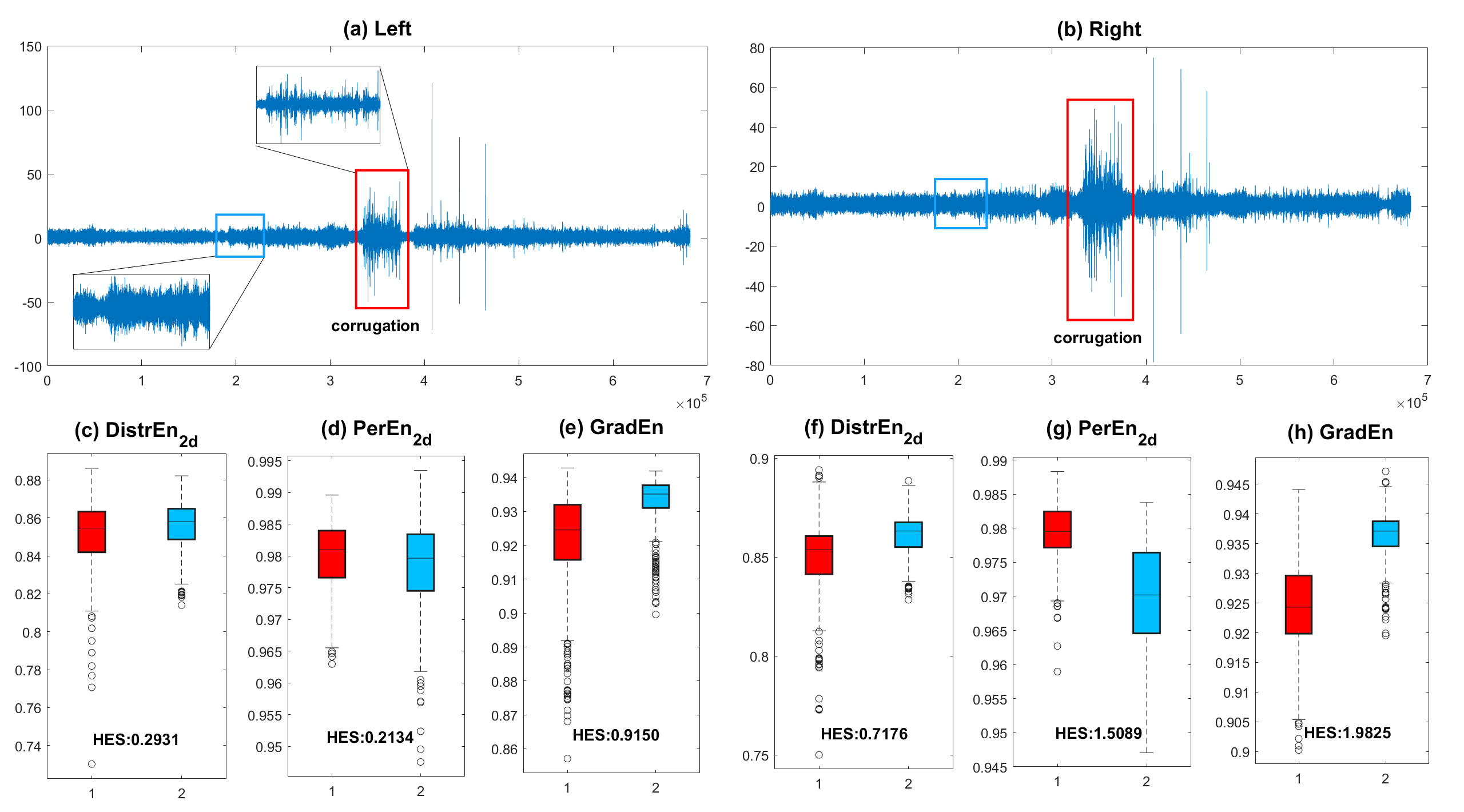}
	\caption{\label{fig15} The boxplot results for distinguishing railway corrugation and normal sections. (a) Left axes signal. (b) Right axes signal. (c) $DistrEn_{2d}$ (left). (d) $PerEn_{2d}$ (left). (e) $GradEn$ (left). (f) $DistrEn_{2d}$ (right). (g) $PerEn_{2d}$ (right). (h) $GradEn$ (right).}
\end{figure}

The detection and diagnosis of railway corrugation is one of the most important technology part for keeping the security of train operation (\cite{railway1,railway2,railway3}). In this section, the acceleration signals collected from the sections where corrugation exists of heavy haul railway are applied to verify the detection ability of $GradEn$. The signals in this experiment are provided by School of Automation and Intelligence, Beijing Jiaotong University. Fig.~\ref{fig15}(a) and Fig.~\ref{fig15}(b) gives the two acceleration signals collected from the left and right axes of the railway section. The corrugation sections are marked by red frame in the picture. In our experiment, the signals with the length of $N=4000$ in both corrugation section (red frame) and normal sections (blue frame) are selected. The sliding window with $N=150$ with each time moves 10 points from the beginning to the end of the time series section. Thus, totally 385 samples of sections from both corrugation and normal signals are obtained. Then, all signals are transformed to distance matrix with embedding dimension $m=3$. The boxplot results of three entropy methods ($DistrEn_{2d}$, $PerEn_{2d}$, $GradEn$) for distinguishing corrugation and normal signals in both left and right axes are shown in Fig.~\ref{fig15}(c-e) and (f-h). Moreover, the coresponding hedge effect size are calculated and given in each picture. The figures and numerical values manifest that $GradEn$ holds the best distinguishing results in both left and right axes signals, which proves that $GradEn$ method provides an effective tool for railway diagnosis research.     

\section{Conclusion}
\label{sec5}
In this study, we propose the $GradEn$ method for image data analysis, building on the 1D SloE algorithm. $GradEn$ offers a novel approach by not only calculating the symbolic pattern but also incorporating amplitude information when extracting features from 2D images. We first discuss the threshold parameters using 2D white noise, followed by experiments with simulated data, including various 2D colored noise types, robust analysis across different lengths of 2D white noise, 2D mixed processes with added noise, and the logistic map. Additionally, we evaluate the computational cost. The results demonstrate the robustness and discriminative power of $GradEn$ in images with diverse characteristics. Real-world datasets, including two texture datasets, a gear fault dataset, and railway corrugation signals, were selected to validate the effectiveness of $GradEn$ in differentiating data under various real-world conditions. The results confirm that $GradEn$ outperforms previous 2D entropy methods in characterizing and distinguishing image data. Therefore, $GradEn$ represents a novel tool for image analysis, offering valuable insights for future research in this field.

Looking ahead, we aim to extend $GradEn$ into a generalized form by exploring additional parameter selections, such as pixel block size and the intervals between points. Furthermore, both the SloE and $GradEn$ algorithms hold promise for expansion into high-dimensional spaces, enabling deeper feature extraction from more complex perspectives.

\section{Statement}
	The paper is under consideration at Pattern Recognition Letters.









\printcredits

\bibliographystyle{model1-num-names}

\bibliography{ref.bib}

\bio{}
\endbio

\endbio

\end{document}